\documentclass[12pt,english]{article}
\usepackage{mathpazo}
\usepackage[T1]{fontenc}
\usepackage[latin9]{inputenc}
\usepackage{geometry}
\usepackage{babel}

\usepackage{graphicx}
\usepackage[unicode=true, pdfusetitle,
 bookmarks=true,bookmarksnumbered=false,bookmarksopen=false,
 breaklinks=false,pdfborder={0 0 1},backref=false,colorlinks=false]
 {hyperref}

\providecommand{\tabularnewline}{\\}

\begin{document}

\title{A Refined Tully-Fisher Relationship and a New Scaling Law for Galaxy
Discs.}
\maketitle
\begin{abstract}
We show how the hypothesis that \emph{galaxy discs conform to self-similar
dynamics} leads to the identification of an annular region of the
optical disc which is such that, corresponding to the Tully-Fisher
scaling law defined on the \emph{exterior} annular boundary, there
is a similar scaling law defined on the \emph{interior} annular boundary.
This result is confirmed at the level of statistical certainty over
several large ORC samples. Furthermore, the same analysis provides
insight into the uncertainties associated with the {}``best'' way
of defining $V_{max}$, the rotation velocity used for the Tully-Fisher
scaling law and $R_{max}$, the galaxy radius at which $V_{max}$
is measured. Finally, as a direct consequence, we are led to a refined
Tully-Fisher law which is largely insensitive to the means by which
a galaxy's rotation velocity is defined.
\end{abstract}

\section{Introduction}

\subsection{General comments}

One of the many problems which hinders our understanding of the underlying
primary processes driving spiral galaxies is the fact that such galaxies
are frequently observed to be {}``afflicted'' by one of the following:
\begin{enumerate}
\item ongoing interactions with external objects;
\item manifest signatures of such interactions in the near past;
\item internal inhomogeities generating local perturbations;
\item presence of bars;
\item unusually active central regions,
\end{enumerate}
and so on. The net effect of these various phenomena is to considerably
complexify the task of identifying the irreducible physics and phenomenology
which define the essential nature of the \emph{spiral galaxy.}

\subsection{Structure suggesting self-similar dynamics in annular discs}

Regardless of the foregoing problems, there still exist various (statistically)
strong signatures which appear to be fundamental; these are:
\begin{enumerate}
\item the existence of the Tully-Fisher scaling relationship ($M\approx a_{0}+b_{0}\log V_{max}$);
\item the long-standing recognition (Danver 1942, Kennicutt 1981) that the
spirality of the spiral arms in disc galaxies can be usefully classified
in terms of the logarithmic spiral, $R=R_{0}\exp(b\theta)$, for radial
displacement $R$ and angular displacement $\theta$.
\end{enumerate}
The second of these signatures, in particular, is consistent with
the hypothesis that \emph{the dynamics in spiral discs conform to
laws of self-similarity.} That is, in the annular sub-regions where
spirality is manifest, \[
\frac{V_{rot}}{V_{rad}}=\frac{\dot{V}_{rot}}{\dot{V}_{rad}}=...=constant,\]
from which, immediately, 

\begin{equation}
V_{rot}=Af(R),\,\,\, V_{rad}=Bf(R)\label{eqn0}\end{equation}
for some undetermined function $f(R)$ and constants $(A,B)$. There
are some obvious comments to be made: 
\begin{enumerate}
\item Since spirality is a property restricted to an \emph{annular} region
of galaxy discs, and since the (logarithmic) spirality property is
implied by (\ref{eqn0}), then this latter hypothesized velocity field
must be considered confined to some annular region. The precise quantitative
definition of this annular region is given in \S\ref{Extraction}; 
\item There is a large amount of galaxy rotation data available which makes
a statistical study of $V_{rot}$ perfectly feasible so that, in principle,
it is possible to hypothesize and test specific proposals for $f(R)$; 
\item Until recently, it was thought that there were essentially no systematic
radial flows in galaxy discs - presumably because of the mass-flow
implications - but it is becoming increasingly recognized that radial
flows are a common feature of galaxy discs so that, in principle at
least, we can conceive the possibility $V_{rad}\propto V_{rot}$ for
the typical galaxy disc. However, the very small amount of radial
flow data presently available renders any large scale statistical
study of radial flows impossible, for some years at least.
\end{enumerate}

\subsection{An hypothesis for $f(R)$ and the manner of its testing}

The topic of how best to express rotation curves in some generic functional
form has received much attention over the years; but all of these
efforts have been focussed on the \emph{whole }rotation curve\emph{.}
By contrast, here we are concerned only with the rotation curve in
a limited annular region (to be precisely defined in \S\ref{Extraction})
of the optical domain where self-similar behaviour appears to be manifested
- that is, we are not concerned with rotation curve behaviour in the
interior regions, and nor are we concerned with the behaviour where
they tend to become (approximately) flat. The most simple possibility
that has any chance of accommodating the variation between discs for
the two-dimensional velocity field \emph{in the optical annulus only}
is $f(R)\equiv R^{\alpha}$, so that our hypothesis becomes \[
V_{rot}\approx AR^{\alpha},\,\,\, V_{rad}\approx BR^{\alpha},\,\,\, R_{min}\leq R\leq R_{max}\]
where $R_{min}$ and $R_{max}$ defined the boundaries of the annular
region to be defined in \S\ref{Extraction}. However, as we have
already indicated, it is only feasible to consider whether or not
$V_{rot}\approx AR^{\alpha}$ at the moment. We assess the validity
of the power-law hypothesis in qualitative and quantitative analyses
and briefly describe these in the immediately following.

\subsubsection{Qualitative assessment}

The adoption of the power-law hypothesis $V_{rot}=AR^{\alpha}$ for
the annular disc leads directly to the following results:
\begin{enumerate}
\item the classical Tully-Fisher scaling relationship,\[
M_{TF}\approx a_{0}+b_{0}\log V_{max},\]
defined on the exterior boundary of the annular disc, emerges automatically
from the correlation which is found to exist between the power-law
parameters $\left(A,\alpha\right)$ and global luminosity properties;
\item the power-law exponent, $\alpha$, which can be estimated independently
of distance scales being determined, allows the definition of an enhanced
Tully-Fisher relation of the type\[
M_{TF}=a+(b+\alpha c)\log V_{max}\,,\]
where $V_{max}$ can be defined in many ways without appreciably affecting
the quality of the resulting $M_{TF}$ determinations;
\item corresponding to the Tully-Fisher scaling relationship on the exterior
boundary of the annular disc, there is a directly analogous scaling
relationship\[
M_{TF}\approx a_{1}+b_{1}\log V_{min}\]
defined on the interior boundary of the annular disc. This result
is entirely new, and lead to the identification of the \emph{dynamically
coherent annular disc} as an objective reality.
\end{enumerate}
The fact that these quantitative results arise directly from the hypothesis
of $V_{rot}=AR^{\alpha}$ for the annular disc provides considerable
circumstantial support for the hypothesis.

\subsubsection{Quantitative assessment}

Generally speaking, when considering power-law fits to data, it is
not possible to say anything more than the power-law does/does not
provide a good fit. In the present case, for various reasons, we can
go much further: we are able to say that, for all practical purposes,
the data behaves \emph{as if} $V_{rot}=AR^{\alpha}$ is the fundamental
law and not merely just a good fit. \\
\\
This is done in the following way: we note that any\emph{ whole-disc}
rotation curve can be expressed as\begin{equation}
\frac{V_{rot}}{V_{0}}=\left(\frac{R}{R_{0}}\right)^{\alpha}+g\left(\frac{R}{R_{0}}\right),\label{eqn0A}\end{equation}
 for arbitrary values $\left(R_{0}>0,V_{0}>0,\,\alpha\right)$ and
for some suitably chosen function, $g(R/R_{0})$. We are able to show
that for each ORC in the large samples considered, values of $\left(R_{0},V_{0},\alpha\right)$
can be chosen (uniquely for each disc) such that\[
\log\left[1+\left(\frac{R_{0}}{R}\right)^{\alpha}g\left(\frac{R}{R_{0}}\right)\right]\approx N(\bar{x},\sigma)\,\,\,{\rm where}\,\,\bar{x}\approx0,\,\,\sigma<<1,\]
over the annular discs of the galaxies concerned. In other words,
$g(R/R_{0})$ itself behaves similarly to a log-normal random variable
with a mean close to zero. For example, for the case of the Mathewson,
Ford and Bucchorn (1992) sample of 900 ORCs, we find $N(\bar{x},\sigma)\approx N(0.002,\,0.07)$.
\\
\\
The implication of this latter result is that, over \emph{large}
samples, the rotational dynamics on the \emph{annular disc} (to be
given an objective operational definition) are described to high precision
in a least-square sense by $V\approx\left[V\right]_{model}\equiv AR^{\alpha}$,
where $(A,\alpha)$ are parameters determined by the ORC in the annular
disc. Alternatively, we can say that deviations from simple power
law behaviour in the annular disc are normally distributed with a
mean close to zero.

\section{The rotation curve samples}

The optical rotation curve samples considered in this analysis are
those of: 
\begin{enumerate}
\item Mathewson, Ford \& Bucchorn (1992) which consists of 900 objects hereafter
referred to as MFB; 
\item Mathewson and Ford (1996) which consists of 1200 objects, hereafter
referred to as MF;
\item Courteaux (1997) which consists of 305 objects, hereafter referred
to as SC; 
\item A composite sample collected from Dale, Giovanelli and Haynes (1997,
1998, 1999) and Dale \& Uson (2000) consisting of 497 objects, hereafter
referred to as DGHU. This sample was provided by permission of Giovanelli
and Haynes.
\end{enumerate}

\subsection{The folding of rotation curves\label{sub:The-folding-of}}

The analysis to be described is predicated upon the ability to fold
rotation curves with high precision and, in this, Persic, Salucci
\& Stel (Persic, M., Salucci, P., \& Stel, F. 1996, MNRAS, 281, 27)
laid out the basic approach to be adopted.\\
\\
Traditionally, the `quick and dirty` approach to folding involves
simply identifying the photometric centre of a disc (an unambiguous
process) and then subtracting the observed redshift at that point
from the whole rotation curve. This approach is perfectly acceptable
if one is primarily interested in the bulk-flow of galaxies. But,
if ones primary interest is in the internal dynamics of the galaxies
themselves (as was Persic et al's), then it is necessary to recognize
that the centre of rotation of a galaxy (its \emph{kinematic centre)}
does not necessarily coincide with its \emph{photometric} centre.
Persic et al found that the best results were obtained if the centre
of folding (the kinematic centre) and the systematic recession velocity
were treated as two free parameters, to be determined such that the
`folded arms were maximally matched`.\\
\\
The difficulty, of course, is in deciding what is meant by `maximally
matched` - especially given that rotation curves are, more often than
not, unevenly sampled in some way; for example, the measurements may
not extend equally on either side of the disc, or measurements may
be taken at unequal intervals on either side of the disc, etc. Persic
et al adopted a labour-intensive, non-automated and qualitative `eye-ball`
approach to the problem, whilst Catinella, Haynes and Giovanelli (AJ,
130, 1037­1048, 2005) have constructed a method based upon using the
two parameters to optimize the fit of Persic et al's \emph{Universal
Rotation Curve} function to their rotation curves. \\
\\
The automatic algorithm developed by this author (A\&A Supp, 140,
247-260, 1999) is also based upon Persic et al's two-parameter method
and, briefly, operates as follows:
\begin{quote}
The two parameters are varied such that, in a Fourier decomposition
of the whole rotation curve (which is purely asymmetric in the ideal
case), a normalized measure of the residual symmetric components is
minimized.
\end{quote}
Due to various problems, such as asymmetric sampling or how many Fourier
modes to use for any given ORC etc, the actual implementation of this
simple algorithm is complicated, but the code is available from this
author. \\
\\
Finally, one very important point, identified by Persic et al,
is that, given a whole set of velocity measurements over a rotation
curve it is necessary to also have quantitative estimates of the absolute
accuracy of each individual velocity measurement available - and then
to use this information as a means of filtering out only the best
individual measurements for the folding process. Broadly speaking,
any individual velocity measurement is \emph{retained} only if its
estimated absolute error $\leq5\%$. For the MFB, MF, SC and DGHU
samples, this requirement led to losses of $35\%$, $25\%$, $46\%$
and $46\%$ respectively of all individual velocity measurements.
The net effect of these data losses meant that many ORCs were then
left with insufficient data points on them to permit a reliable folding.
The overall attrition rate of ORCs lost to the overall analysis via
this process were $3\%$, $4\%$, $7\%$ and $8\%$ respectively.

\section{The annular disc: A dynamically coherent whole\label{sec: annular-disc:}}

The most significant result of this section is the demonstration that
the \emph{annular disc} is a dynamically coherent and objectively
defined component of the total disc. Apart from giving the algorithmic
definition of the annular disc, the demonstration consists in showing
how the \emph{interior} boundary of the annular disc is a dynamical
transition region between the inner and outer parts of the disc on
which a scaling law, directly analogous to the classical Tully-Fisher
scaling law, is defined.

\subsection{Extraction of the annular disc from the whole disc\label{Extraction}}

The process to be described is based on the hypothesis that $V\approx\left[V\right]_{model}\equiv AR^{\alpha}$
describes rotation velocities in ORCs in certain annular regions which
are to be extracted from the whole disc. Before describing the extraction
algorithm, there are two important notes to be made:
\begin{enumerate}
\item By the `whole disc` in this context, we mean the complete set of velocity
data for the disc with the exception of any filtered-out poor-quality
individual measurements (cf last section). The annular disc is extracted
from the whole disc using a statistical algorithm based upon the technique
of linear regression: following conventional definitions, an observation
is reckoned to be \emph{unusual} if the predictor is unusual, or if
the response is unusual. For a $p$-parameter model, a predictor is
commonly defined to be unusual if its leverage $>3p/N$, when there
are $N$ observations. In the present case, we have a two-parameter
model so that $p=2$. Similarly, the \emph{response} is commonly defined
to be unusual if its standardized residual $>2$.
\item The prescription $V\approx AR^{\alpha}$ can only be completely specified
for any given ORC when the distance scale has been set. However, the
exponent $\alpha$ is independent of the distance scale; consequently,
the identification of any annular region over which the approximate
power-law may be valid is also independent of the distance scale.
For this reason, the extraction algorithm described below assumes
that radial displacements are specified in radians, $R_{(radians)}$
say; the corresponding calculated $A$ is denoted as $A^{*}$. 
\end{enumerate}
The computation of $(\alpha,\ln A^{*})$, for any given folded and
inclination-corrected rotation curve, can now be described using the
following algorithm: 
\begin{enumerate}
\item Assume the model $V\approx\left[V\right]_{model}=A^{*}R_{(radians)}^{\alpha}$
for the rotation velocity in some annular region of the disc, to be
determined. But initially, use the whole disc ORC;
\item Form an estimate of the parameter-pair $(\alpha,\ln A^{*})$ by regressing
the $\ln V$ data on the $\ln R_{(radians)}$ data for the folded
ORC; 
\item Determine if the \emph{innermost} observation only is an \emph{unusual}
observation in the sense defined above; 
\item If the innermost observation is unusual, then exclude it from the
computation and repeat the process from (2) above on the reduced data-set; 
\item If the innermost observation is \emph{not} unusual, then no further
computation is required:

\begin{enumerate}
\item the remaining set of points at this stage are considered to define
the extracted annular region. The minimum value of $R$ for any given
disc, $R_{min}$ say, defines the radius of the interior boundary
of the extracted annulus and $V_{min}$ denotes the rotation velocity
on this interior boundary;
\item the current value of $\alpha$ defines the exponent of the power-law
over the extracted annular region;
\item the value $A^{*}$ has no particular significance since a distance
scale has yet to be set. 
\end{enumerate}
\end{enumerate}
This algorithm has the result that, on average, $(\alpha,\ln A^{*})$
is computed on the exterior 88\% of the data points in each ORC of
the MFB sample, the exterior 87\% of data points in each ORC of the
MF sample, the exterior 91\% of data points in each ORC of the DGHU
sample and the exterior 91\% of the data points in each ORC of the
SC sample.

\subsection{An old result quantified and refreshed }

\begin{figure}
\includegraphics[scale=0.5]{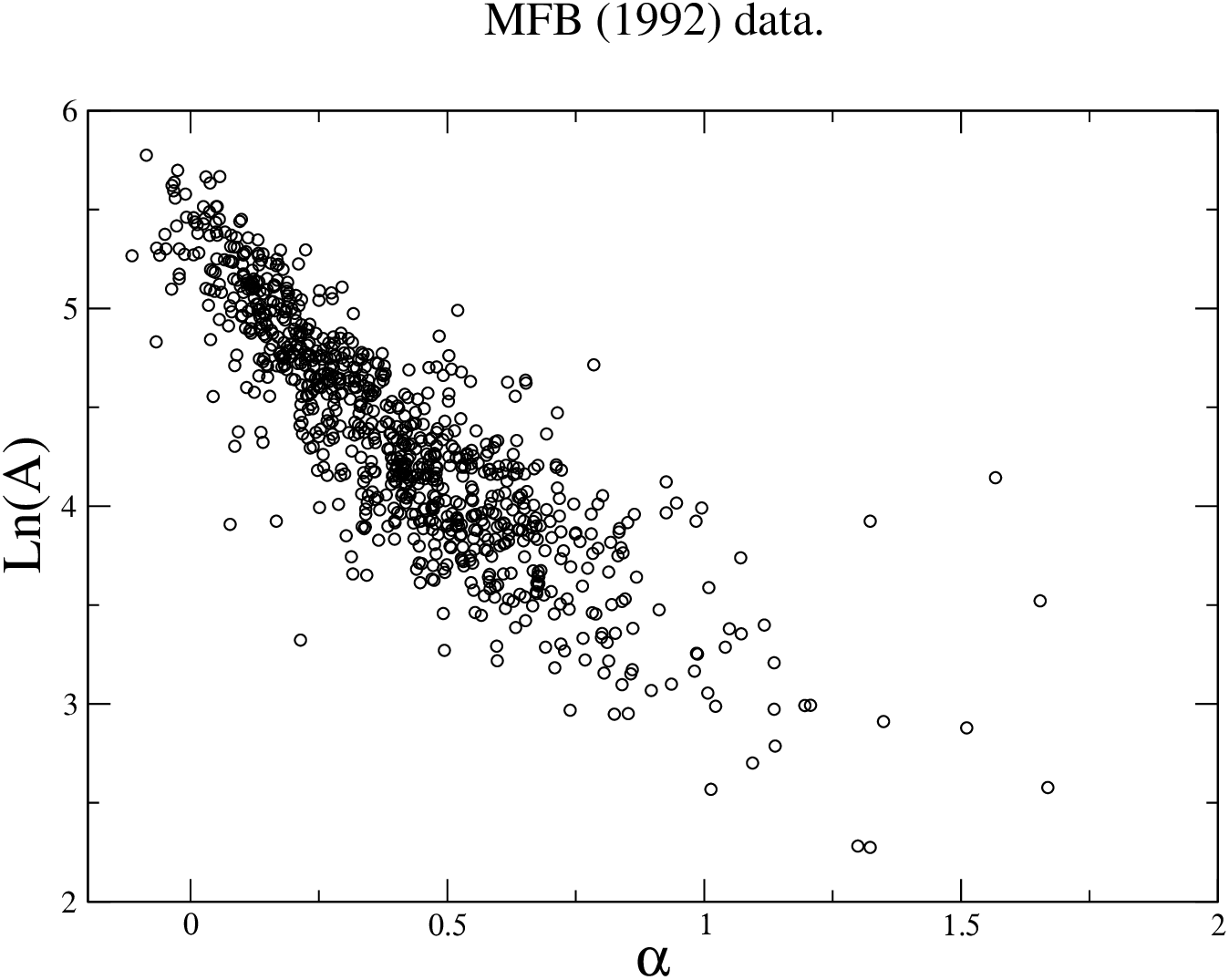}

\caption{\label{cap: Figure 1}}

\end{figure}
 It has been known for a long time that, in a qualitative sense, the
steeper the rise of a rotation curve on its interior portion, then
the flatter the rotation curve is in its exterior portion. \\
\\
Given our hypothesis that $V\approx AR^{\alpha}$ on the annular
disc, then it is clear that the parameter $A$ has the \emph{numerical
value} of the approximate rotation velocity at 1$kpc$ (given that
$R$ is in units of $kpc$) - as estimated by using the power law
to extrapolate (or interpolate) the data to 1$kpc$. It follows that
the value of $A$ can also be considered as a proxy measure for the
steepness of the initial rise of the rotation curve in which case,
given the qualitative statement above, we should expect $A$ and $\alpha$
to be in an inverse relation to each other. \\
\\
For the sake of demonstrating the reality of this relationship,
we use MFB's Tully-Fisher distances to set the radial distance scale
so that $A$ can be computed for each galaxy in sample. Figure \ref{cap: Figure 1}
then shows the scatter diagram plotting $(\alpha,\ln A)$, computed
as in \S\ref{Extraction} for each foldable galaxy in the Mathewson
et al (1992) sample. The powerful nature of their inverse relation
is clearly apparent. \\
\\
To summarize, the given analysis quantifies old qualitative knowledge
by providing an explicit and well-defined partition of the optical
disc into distinct dynamical regions: the interior disc containing
the steep initial rise, and the exterior annular disc over which -
as we shall demonstrate - the rotational dynamics are described to
extremely high fidelity by the $V\approx AR^{\alpha}$.

\subsection{Scaling laws and the annular disc}

The algorithm described in \S\ref{Extraction} will always produce
a result in the form of an extracted annular disc. So, the significant
questions are: 
\begin{enumerate}
\item Does the process described have any basis in physics so that the annular
disc can be identified as a physically coherent distinct sub-component
of the whole disc \emph{?}
\item Or is the data reduction process essentially arbitrary, the product
of which has no physical significance whatsoever?
\end{enumerate}
In the following, we show that a scaling law similar to the Tully-Fisher
law which applies at the exterior boundary of the annular disc also
applies on its interior boundary, $R\equiv R_{min}$ determined in
\S\ref{Extraction},  and at similar levels of statistical significance.
Thus, this interior boundary is a boundary of objective physical significance
which scales according to luminosity and which defines, in effect,
the transition within the disc between one form of behaviour and another.

\subsubsection{The interior Tully-Fisher law}

In the following, we establish the existence of a Tully-Fisher-type
law on the interior boundary of the annular disc. The results of the
individual regressions are given below:\begin{eqnarray}
{\rm MFB:}\,\,\,\, M_{TF} & \approx & (-14.60\pm0.15)+(-1.40\pm0.03)\ln V_{min}\,,\nonumber \\
 &  & n=849,\,\,\, t_{grad}=-42,\,\, R_{adj}^{2}=67\%\,;\nonumber \\
\nonumber \\{\rm MF:}\,\,\,\, M_{TF} & \approx & (-14.79\pm0.13)+(-1.36\pm0.03)\ln V_{min}\,,\nonumber \\
 &  & n=1070,\,\, t_{grad}=-45,\,\, R_{adj}^{2}=65\%\,;\nonumber \\
\label{eqn2}\\{\rm SC:}\,\,\,\, M_{TF} & \approx & (-14.89\pm0.26)+(-1.36\pm0.05)\ln V_{min}\,,\nonumber \\
 &  & n=275,\,\,\,\, t_{grad}=-25,\,\, R_{adj}^{2}=69\%\,;\nonumber \\
\nonumber \\{\rm DGHU:}\,\,\,\, M_{TF} & \approx & (-11.81\,\pm0.24)+(-1.95\pm0.05)\ln V_{min}\,,\nonumber \\
 &  & n=487,\,\,\,\, t_{grad}=-38,\,\, R_{adj}^{2}=75\%\,.\nonumber \end{eqnarray}
The quoted $t$-statistic is each case refers to the estimated gradient
value and it is quite obvious that, in every case, the dependence
of the Tully-Fisher estimated absolute luminosity, $M_{TF}$, upon
the rotational velocity on the interior boundary of the annular disc
is established at the level of statistical certainty. \\
\\
It is interesting to note the remarkable consistency in the regression
models arising from the MFB, MF and SC samples, with that arising
from the DGHU sample being significantly different. the reasons for
the discrepant nature of the DGHU results are not clear.

\subsection{Conclusions for the annular disc}

It has been demonstrated, at the level of statistical certainty, that
a scaling law analogous to that of Tully-Fisher, exists on the interior
boundary on the extracted annular disc. This establishes that: 
\begin{enumerate}
\item the extraction process is not arbitrary, but gives a result which
has an objective physical significance; 
\item the annular disc is a physically coherent distinct component of the
complete disc;
\item the boundary between the inner-disc and the annular disc is a physical
transition boundary between one form of dynamical behaviour and another;
\item the power-law hypothesis, $V=AR^{\alpha}$, for the annular disc provides
a good statistical description for the rotational dynamics in annular
discs over large samples. This conclusion will be quantified in a
later section.
\end{enumerate}

\section{An enhanced Tully-Fisher law: the parameter $\alpha$\label{sec:4} }

The classical Tully-Fisher scaling law is given by\[
M_{TF}=a+b\log V_{rot}\]
where $V_{rot}$ is an estimated \emph{maximal} (or characteristic)
rotation velocity. The actual precise definition of $V_{rot}$ is
highly problematical - especially when working in the optical - and
very much effort has been expended in trying to construct algorithmic
definitions for it. In this section, we consider an \emph{enhanced}
Tully-Fisher scaling law\begin{equation}
M_{TF}=a+\left(b+\alpha c\right)\log V_{rot}\,,\label{eqn4}\end{equation}
where the parameter $\alpha$ is the exponent in the power-law fit
$V_{rot}=AR^{\alpha}$ to any given rotation curve, and we show that
the inclusion of the term $\alpha\log V_{rot}$ in (\ref{eqn4}): 
\begin{itemize}
\item makes the performance of the Tully-Fisher scaling law insensitive
to the definition of $V_{rot}$;
\item improves the performance of the Tully-Fisher scaling law when the
chosen definition of $V_{rot}$ is sub-optimal.
\end{itemize}

\subsection{$V_{rot}$ definitions used in the present analysis }

The various definitions of $V_{rot}$ used in this analysis to make
the point are listed as follows: 
\begin{itemize}
\item The \emph{histogram method,} for which $V_{rot}\equiv V_{hist}\equiv0.5\left(V_{N}-V_{100-N}\right)$
where $V_{N}$ is the $N$-th percentile velocity. Versions of this
are used by MFB, MF and DGHU;
\item The \emph{model linewidth method,} for which $V_{rot}=V_{model}$
where $V_{model}$ is the interpolated velocity at some formally defined
position on some phenomenologically defined function fitted to the
rotation curve. SC investigated a variety of these to arrive at his
$V_{max}$ (which is the peak velocity on his model curve) and his
$V_{2.2}$ (which is his model velocity at $R=2.15\times$(disc scale
length)). Catinella et al (2005) use their \emph{polyex} function
to arrive at a similar definition for $V_{model}$.
\item Various authors (eg Persic \& Salucci 1995) have suggested using $V_{rot}=V_{opt}$
where $V_{opt}$ is the (usually interpolated) rotation velocity at
the optical radius, defined as $R_{opt}\equiv R_{83}$. In the following
analysis, $R_{opt(radians)}$ is determined directly from the photometry
and we estimate $V_{opt}$ by fitting $V_{rot}=A^{*}R_{radians}^{\alpha}$
to the ORC data prior to the distance scale being set ($\alpha$ is
independent of the distance scale) and then defining $V_{opt}=A^{*}R_{opt(radians)}^{\alpha}$.
\end{itemize}

\subsection{The analysis}

For each of the four samples, MFB, MF, DGHU and SC, available to us
we:
\begin{itemize}
\item calibrate the classical Tully-Fisher scaling law using Hubble distances
together with:

\begin{itemize}
\item the rotational velocities provided by the authors concerned. For MFB,
MF and DGHU, these are estimated using versions of the histogram method
so that $V_{rot}=V_{hist}$. For SC, these are estimated using various
model linewidths; we consider just two of these, his $V_{2.2}$ and
$V_{max}$;
\item the rotation velocities interpolated to the optical radii, $R_{83}$,
$V_{opt}$ say;
\end{itemize}
\item calibrate the enhanced Tully-Fisher scaling law, (\ref{eqn4}), for
the same samples and definitions of $V_{rot}$; 
\item compare the fits of the classical Tully-Fisher scaling laws to the
enhanced scaling laws over each of the four samples and for each of
the definitions of $V_{rot}$ considered.
\end{itemize}
The detailed analyses are given in the appendices but the results
are summarized in figure \ref{fig3} : %
\begin{figure}
\includegraphics[scale=0.6]{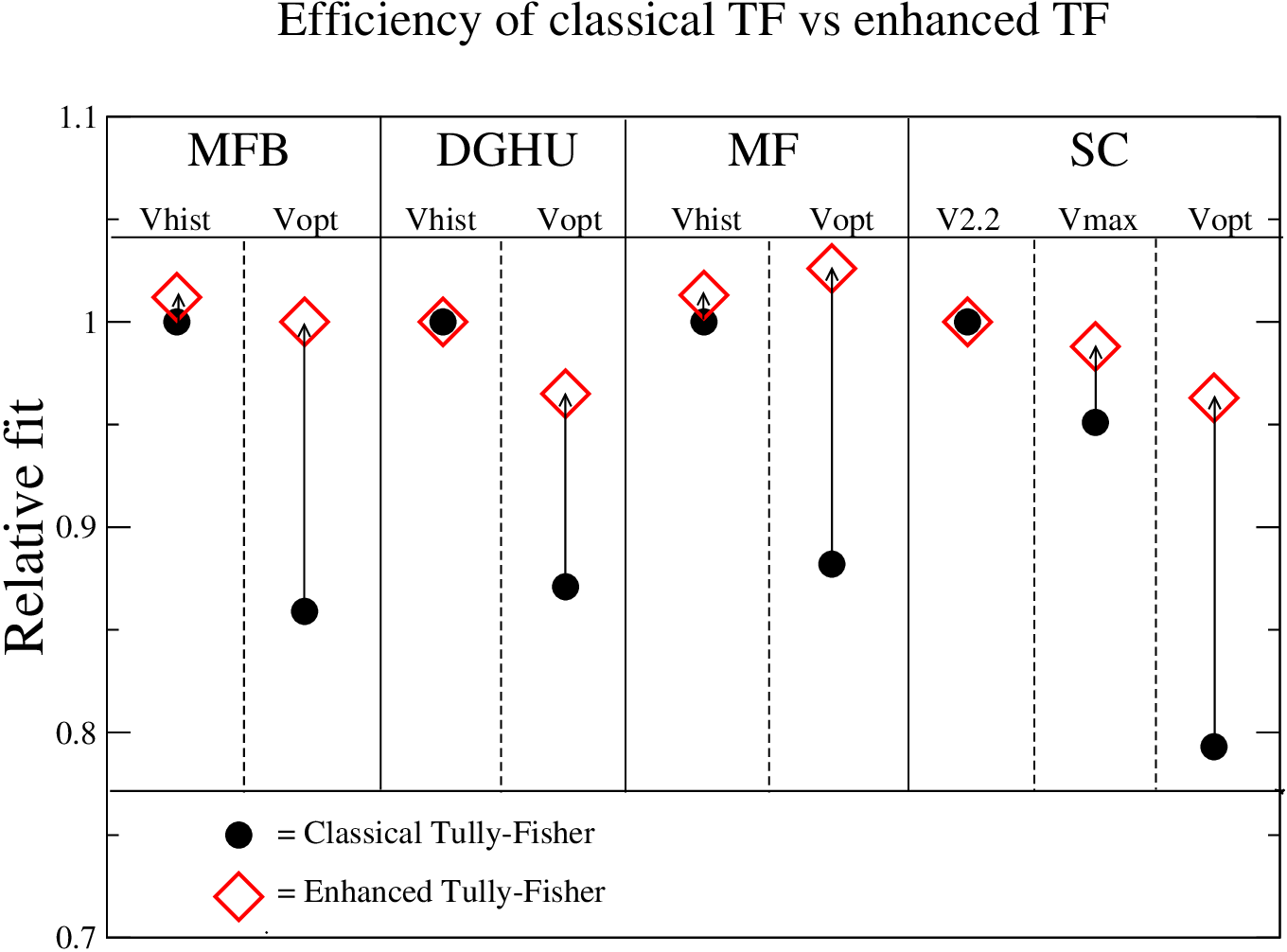}

\caption{\label{fig3}}

\end{figure}
The samples are given in the major columns and the major columns are
subdivided according to which definition of $V_{rot}$ is used. The
relative scaling-law performance is estimated using a normalized measure
of the fit of the scaling law to the data, as follows:
\begin{itemize}
\item for each sample, the classical scaling law is calibrated using $V_{hist}$
in the case of MFB, MF and DGHU and $V_{2.2}$ in the case of SC;
\item in each case, an estimate of fit to the data is given by the adjusted
$R^{2}$ parameter, labelled $R_{MFB}^{2}$, $R_{MF}^{2}$, $R_{DGHU}^{2}$
and $R_{SC}^{2}$ respectively;
\item each of $R_{MFB}^{2}$, $R_{MF}^{2}$, $R_{DGHU}^{2}$ and $R_{SC}^{2}$
is normalized to unity;
\item for every other calibration of either the classical scaling law or
the enhanced scaling law, the corresponding estimate of fit, $R^{2}$,
is normalized using one of $R_{MFB}^{2}$, $R_{MF}^{2}$, $R_{DGHU}^{2}$
and $R_{SC}^{2}$, as appropriate. 
\end{itemize}
In figure \ref{fig3}, the filled circles respresent the normalized
$R^{2}$ values arising from the classical scaling law whilst the
open diamonds represent these values arising from the enhanced scaling
law. The figure makes it very clear that the effectiveness of the
enhanced scaling-law is very much less sensitive to the definition
of $V_{rot}$ than that of the classical scaling law. In particular:
\begin{itemize}
\item in every case, $V_{rot}=V_{opt}$ performs poorly in the classical
scaling law but performs comparably with all other choices in the
enhanced scaling law; 
\item for the DGHU and SC samples, the choices $V_{hist}$ and $V_{2.2}$
respectively appear to be optimal in the sense that the use of the
enhanced scaling law makes no measurable difference over the classical
scaling law;
\item for the MFB and MF samples, the enhanced scaling law gives measurable,
but small, improvements over the classical scaling law when $V_{hist}$
is used. The detailed analysis, given in the appendices, shows that
these improvements are strongly significant in a statistical sense;
\item for the SC sample, the enhanced scaling law gives much larger measurable
improvement over the classical scaling law when $V_{2.2}$ is used. 
\end{itemize}

\subsection{Interim conclusions}

We have demonstrated that the inclusion of the predictor $\alpha\log V_{rot}$
has the potential to greatly improve the predictive power of the Tully-Fisher
scaling law but that, in practice, improvements are strongly contingent
of the precise definition of $V_{rot}$ employed. It appears that
there is an optimal definition of $V_{rot}$ which is similar to those
employed by DGHU and SC but, whichever definition is used, suboptimal
choices are brought to near-optimality by the inclusion of $\alpha\log V_{rot}$
as a predictor.

\section{The classical Tully-Fisher scaling law\label{sec:Natural}}

So far, we have shown that new insights into scaling laws for galaxy
discs arise when we assume that $V\approx AR^{\alpha}$ over some
partition of galaxy discs, $R_{min}<R<R_{max}$ say. In the following,
we: 
\begin{enumerate}
\item show how the Tully-Fisher scaling law emerges from the power-law hypothesis
for the annular disc; 
\item clarify why $\alpha\ln V_{rot}$ can be expected to be a significant
predictor according to the precise definition chosen for $V_{rot}$.
\end{enumerate}

\subsection{The power-law hypthesis and the Tully-Fisher scaling law\label{sub: The-TF}}

Given the hypothesis $V=AR^{\alpha}$ for the rotation velocity in
the annular disc, where $(A,\alpha)$ vary between discs, then for
an arbitrarily chosen point $\left(R_{0},V_{0}\right)$ on any given
rotation curve, we have \begin{equation}
\ln A=\ln V_{0}-\alpha\ln R_{0}.\label{eqn6b}\end{equation}
But we know that if, for example, we choose $V_{0}=V_{rot}$ on any
given ORC then correlations of the type $V_{0}\equiv V_{rot}=f(M,\alpha)$
exist. \\
\\
In practice, an exploration of the MFB, MF, DGHU and SC samples
(see \S\ref{sec:The-detailed} for the quantitative numerical details)
shows that models of the general type\begin{equation}
\ln A\approx\left[\ln A\right]_{model}\equiv a_{0}+a_{1}M+\alpha\left(b_{1}+b_{2}M+b_{3}\ln S\right),\label{eqn6c}\end{equation}
where $S$ is the surface brightness, are comprehensive (that is,
all significant predictors are included) and account for, typically,
95\% of the variation in $\ln A$ data. \\
\\
A quick comparison of (\ref{eqn6b}) with (\ref{eqn6c}) shows
that the decomposition \begin{equation}
\ln V_{0}=a_{0}+a_{1}M,\,\,\,\ln R_{0}=-\left(b_{1}+b_{2}M+b_{3}\ln S\right)\label{eqn6d}\end{equation}
is possible so that, immediately, we have the classical Tully-Fisher
scaling law, together with a corresponding relation giving the value
of $R\equiv R_{0}$ at which $V_{0}$ is measured. However, the decomposition
(\ref{eqn6d}) is \emph{not unique} and in this observation we find
an understanding of the appearance of $\alpha\ln V_{rot}$ as a predictor.

\subsection{The predictor $\alpha\ln V_{rot}$ in the Tully-Fisher scaling law}

The considerations of \S\ref{sub: The-TF} above lead to an understanding
of why $\alpha\ln V_{rot}$ is also a predictor in Tully-Fisher calibrations,
and of why its significance varies according to the definition of
$V_{rot}$ employed.\\
\\
In effect, (\ref{eqn6d}) defines both a particular \emph{characteristic
velocity} on an ORC and the \emph{characteristic radius} at which
this velocity is to be measured, both being determined in terms of
luminosity properties of the galaxy concerned. However, suppose that
we wish to defined the characteristic velocity at\[
\ln\hat{R}_{0}\equiv\ln R_{0}+\Delta\ln R_{0}\]
 rather than at $\ln R_{0}$. Then (\ref{eqn6d}) shows that, since
$M$ and $S$ are fixed for the galaxy, $\Delta\ln R_{0}\neq0$ can
only arise from some perturbation of $(b_{1},b_{2},b_{3})$ - for
example, $\Delta\ln R_{0}=-\Delta b_{2}M$. In this case, (\ref{eqn6b})
can only be satisfied if (\ref{eqn6d}) becomes\[
\ln\hat{V}_{0}=a_{0}+\left(a_{1}-\alpha\Delta b_{2}\right)M,\,\,\,\ln\hat{R}_{0}=-\left(b_{1}+b_{2}M+\Delta b_{2}M+b_{3}\ln S\right).\]
Thus, the simple act of redefining the characteristic radius according
to $R_{0}\rightarrow\hat{R}_{0}$ immediately causes $\alpha\ln\hat{V}_{0}$
to become a predictor for $M$\@. \\
\\
In other words, whether or not $\alpha\ln V_{0}$ is a significant
predictor in any given case depends entirely upon the definition adopted
for $V_{0}$ which, in effect, amounts to systematically selecting
a particular definition for $R_{0}$.

\section{The detailed analysis of the four samples\label{sec:The-detailed}}

In the following, we show how the considerations of the previous section
work in practice in the context of the samples of MFB, MF, SC and
DGHU: in particular, we explore the model

\begin{equation}
\ln A\approx\left[\ln A\right]_{model}\equiv a_{0}+a_{1}M_{TF}+\alpha\left(b_{1}+b_{2}M_{TF}+b_{3}\ln S_{TF}\right)\label{eqn10}\end{equation}
for those samples where, in each case, $M_{TF}$ is estimated using
Tully-Fisher relationships calibrated by the authors concerned.

\subsection{The sample of MFB}

The MFB sample contains 864 ORCs that we were able to fold and, after
removing 28 of the more extreme outliers, the model (\ref{eqn10})
is defined by the table:\\

~~~~~~~~~~~~~~~~~~\begin{tabular}{|l|ccccc|c|}
\hline 
Coeffs & $a_{0}$ & $a_{1}$ & $b_{1}$ & $b_{2}$ & $b_{3}$ & $R_{adj}^{2}$\tabularnewline
\hline
Estimate & -0.946 & -0.286 & 5.590 & 0.475 & 0.586 & $95\%$\tabularnewline
Std Error & 0.149 & 0.007 & 0.286 & 0.015 & 0.020 & \tabularnewline
$t$-statistic & -6 & -41 & 20 & 31 & 29 & \tabularnewline
\hline
\end{tabular}\\
\\
\\
so that, from (\ref{eqn6d}) we immediately get\begin{equation}
\ln V_{0}=-0.946-0.286M_{TF},\,\,\,\,\ln R_{0}=-5.590-0.475M_{TF}-0.586\ln S_{TF}.\label{eqn9A}\end{equation}
The first of these two relations gives, directly (to two decimal places)\[
M_{TF}=-8.05\log_{10}V_{0}-3.31\]
which is very close to MFB's calibration of the Tully-Fisher relation,\[
M_{TF}=-7.96\log_{10}V_{rot}-3.30.\]

\subsection{The sample of MF}

The MF sample contains 1083 ORCs that we were able to fold and, after
removing 26 of the more extreme outliers, the model (\ref{eqn10})
is defined by the table:

~~~~~~~~~~~~~~~~~~~~\begin{tabular}{|l|ccccc|c|}
\hline 
Coeffs & $a_{0}$ & $a_{1}$ & $b_{1}$ & $b_{2}$ & $b_{3}$ & $R_{adj}^{2}$\tabularnewline
\hline
Estimate & -1.199 & -0.298 & 5.976 & 0.469 & 0.477 & $94\%$\tabularnewline
Std Error & 0.141 & 0.007 & 0.288 & 0.015 & 0.019 & \tabularnewline
$t$-statistic & -9 & -45 & 21 & 32 & 25 & \tabularnewline
\hline
\end{tabular}\\
\\
\\
so that, from (\ref{eqn6d}) we immediately get\begin{equation}
\ln V_{0}=-1.199-0.298M_{TF},\,\,\,\,\ln R_{0}=-5.976-0.469M_{TF}-0.477\ln S_{TF}.\label{eqn9B}\end{equation}
The first of these two relations gives, directly (to two decimal places)\[
M_{TF}=-7.45\log_{10}V_{0}-4.62\]
which is to be compared with MF's (MFB's) calibration of the Tully-Fisher
relation,\[
M_{TF}=-7.96\log_{10}V_{rot}-3.30.\]

\subsection{The sample of DGHU}

The DGHU sample contains 497 ORCs that we were able to fold and, after
removing 24 of the more extreme outliers, the model (\ref{eqn10})
is defined by the table:\\

~~~~~~~~~~~~~~~~~~~\begin{tabular}{|l|ccccc|c|}
\hline 
Coeffs & $a_{0}$ & $a_{1}$ & $b_{1}$ & $b_{2}$ & $b_{3}$ & $R_{adj}^{2}$\tabularnewline
\hline
Estimate & -1.412 & -0.307 & 5.540 & 0.434 & 0.415 & $96\%$\tabularnewline
Std Error & 0.180 & 0.008 & 0.420 & 0.021 & 0.024 & \tabularnewline
$t$-statistic & -8 & -37 & 13 & 20 & 17 & \tabularnewline
\hline
\end{tabular}\\
\\
\\
so that, from (\ref{eqn6d}) we immediately get\begin{equation}
\ln V_{0}=-1.412-0.307M_{TF},\,\,\,\,\ln R_{0}=-5.540-0.434M_{TF}-0.415\ln S_{TF}.\label{eqn9C}\end{equation}
The first of these two relations gives, directly (to two decimal places)\[
M_{TF}=-7.50\log_{10}V_{0}-4.60\]
which is to be compared with DGHU's calibration of the Tully-Fisher
relation,\[
M_{TF}=-7.68\log_{10}V_{rot}-4.23.\]

\subsection{The sample of SC using $V_{max}$}

The SC sample contains 282 ORCs that we were able to fold and, after
removing 24 of the more extreme outliers, the model (\ref{eqn10})
is defined by the table:\\

~~~~~~~~~~~~~~~~~~~\begin{tabular}{|l|ccccc|c|}
\hline 
Coeffs & $a_{0}$ & $a_{1}$ & $b_{1}$ & $b_{2}$ & $b_{3}$ & $R_{adj}^{2}$\tabularnewline
\hline
Estimate & -3.225 & -0.394 & 9.282 & 0.604 & 0.288 & $98\%$\tabularnewline
Std Error & 0.219 & 0.010 & 0.744 & 0.0235 & 0.026 & \tabularnewline
$t$-statistic & -15 & -39 & 12 & 17 & 11 & \tabularnewline
\hline
\end{tabular}\\
\\
\\
so that, from (\ref{eqn6d}) we immediately get\begin{equation}
\ln V_{0}=-3.225-0.394M_{TF},\,\,\,\,\ln R_{0}=-9.282-0.604M_{TF}-0.288\ln S_{TF}.\label{eqn9D}\end{equation}
The first of these two relations gives, directly (to two decimal places)\[
M_{TF}=-5.84\log_{10}V_{0}-8.19\]
which is to be compared with SC's calibration of the Tully-Fisher
relation using his $V_{max}$,\[
M_{TF}=-6.19\log_{10}V_{rot}-7.50.\]

\section{A quantitative test of the power-law hypothesis }

So far, the assumption of the power-law hypothesis, $V_{rot}=AR^{\alpha}$,
has led to considerable new insight into the scaling properties of
galaxy discs so that \emph{ipso facto} the hypothesis has great utility
in practice; however, these results amount to \emph{qualitative} evidence
in favour of the hypothesis. \\
\\
In the following we give a \emph{quantitative} data analysis which
shows that, at the level of statistical certainty, rotation velocities
within the annular disc (defined in \S\ref{sec: annular-disc:})
of spirals behave \emph{as if} the power law hypothesis \[
\left(\frac{V}{V_{0}}\right)\approx\left(\frac{R}{R_{0}}\right)^{\alpha},\,\,\, R_{min}<R<R_{max}\,,\]
where $\left(\alpha,V_{0},R_{0}\right)$ are parameters unique to
each galaxy, is the physical law governing rotation velocity within
spiral discs. %
\begin{figure}
~~~~~~~~~~~~\includegraphics[scale=0.35,angle=90]{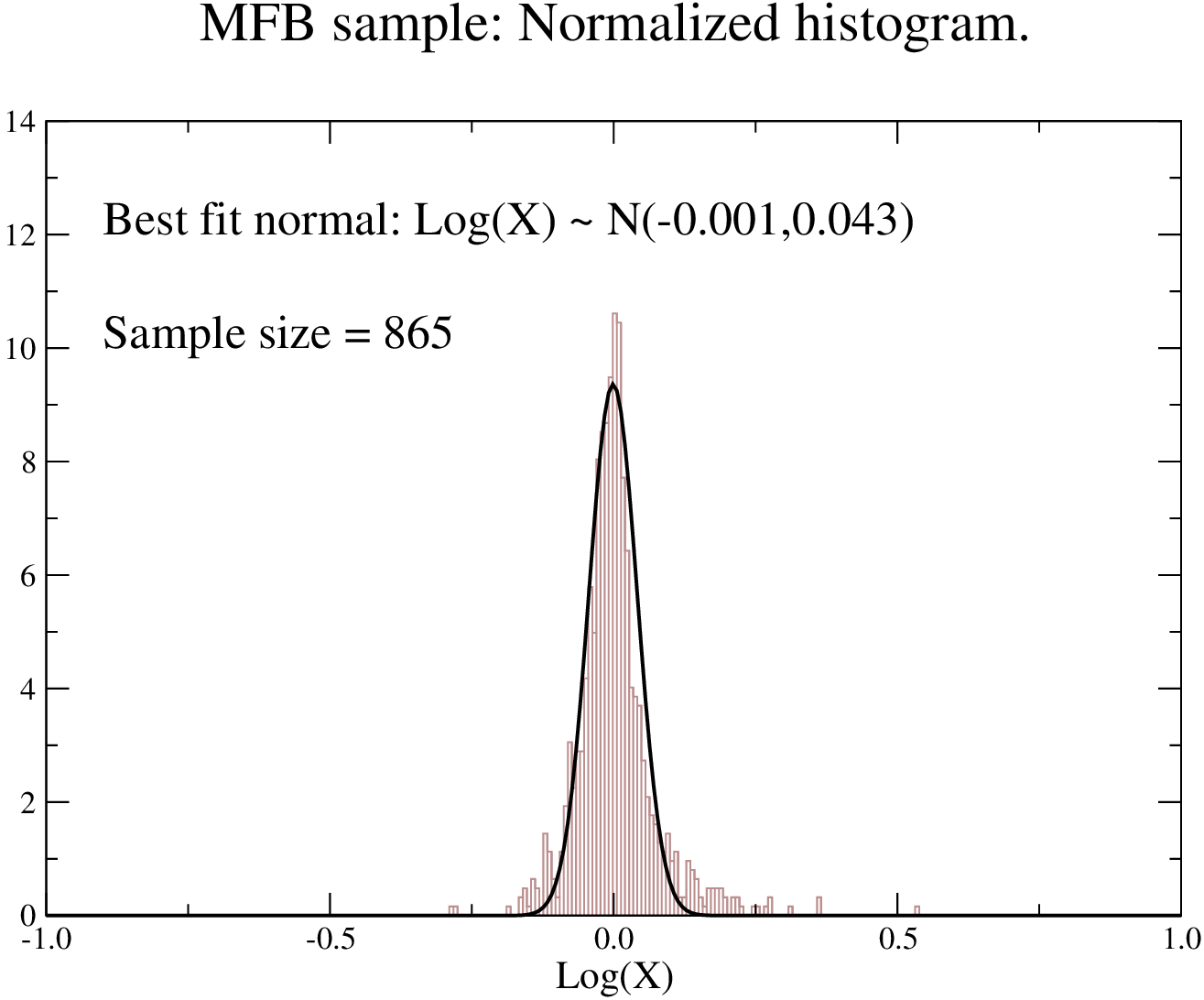}~~~~~~~\includegraphics[scale=0.35,angle=90]{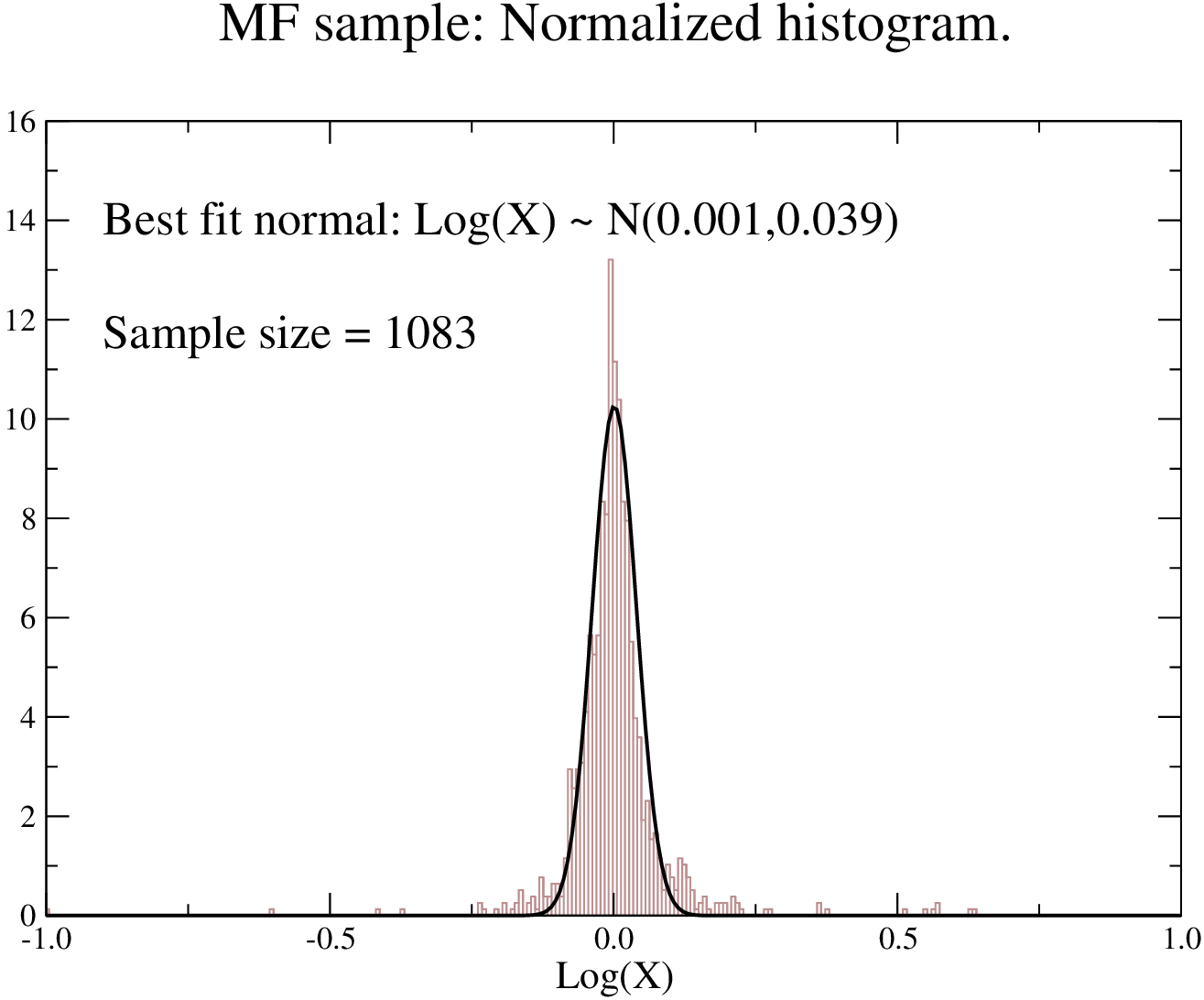}\\
\\
\\
\\

~~~~~~~~~~~~\includegraphics[scale=0.35,angle=90]{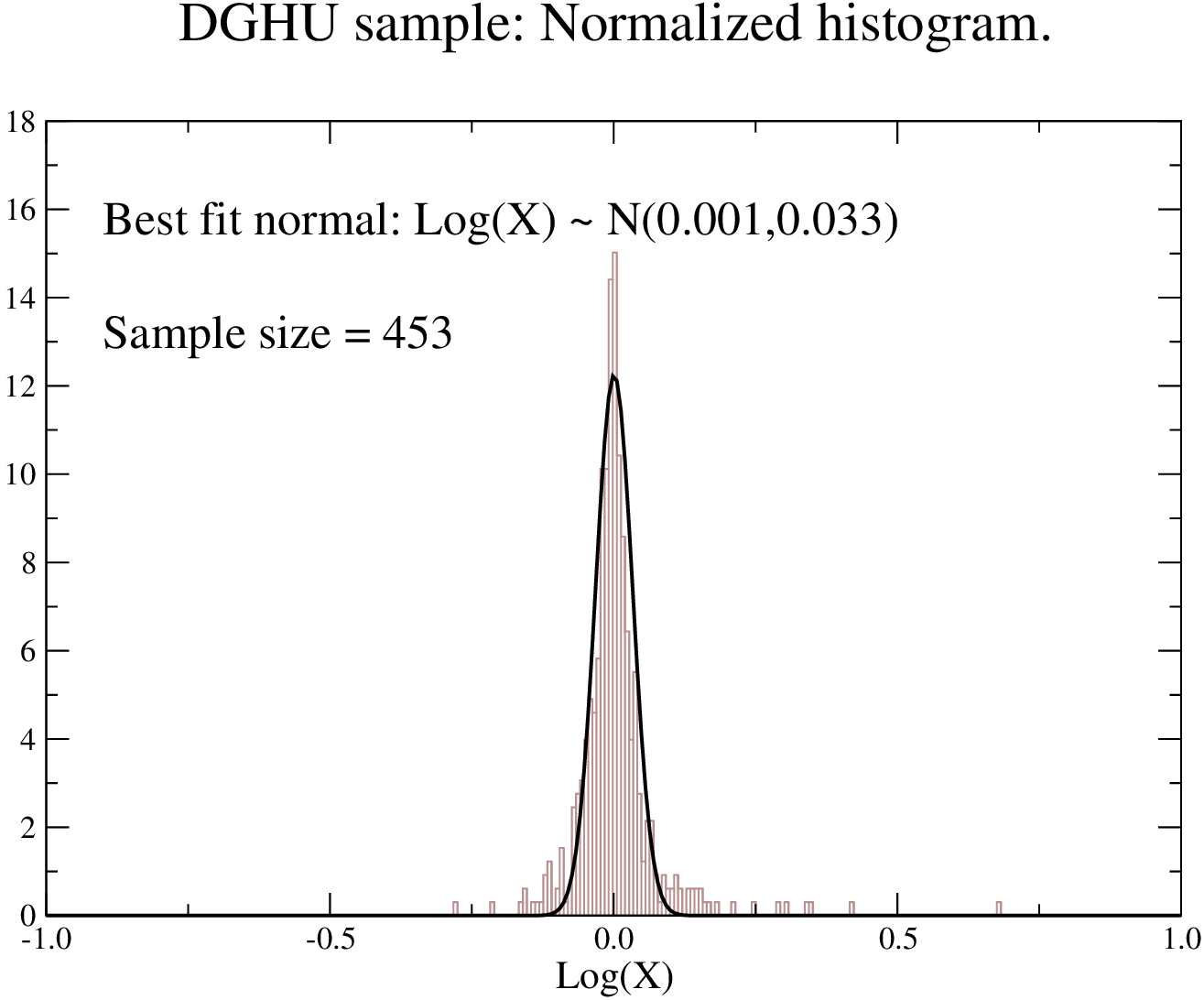}~~~~~~~\includegraphics[scale=0.35,angle=90]{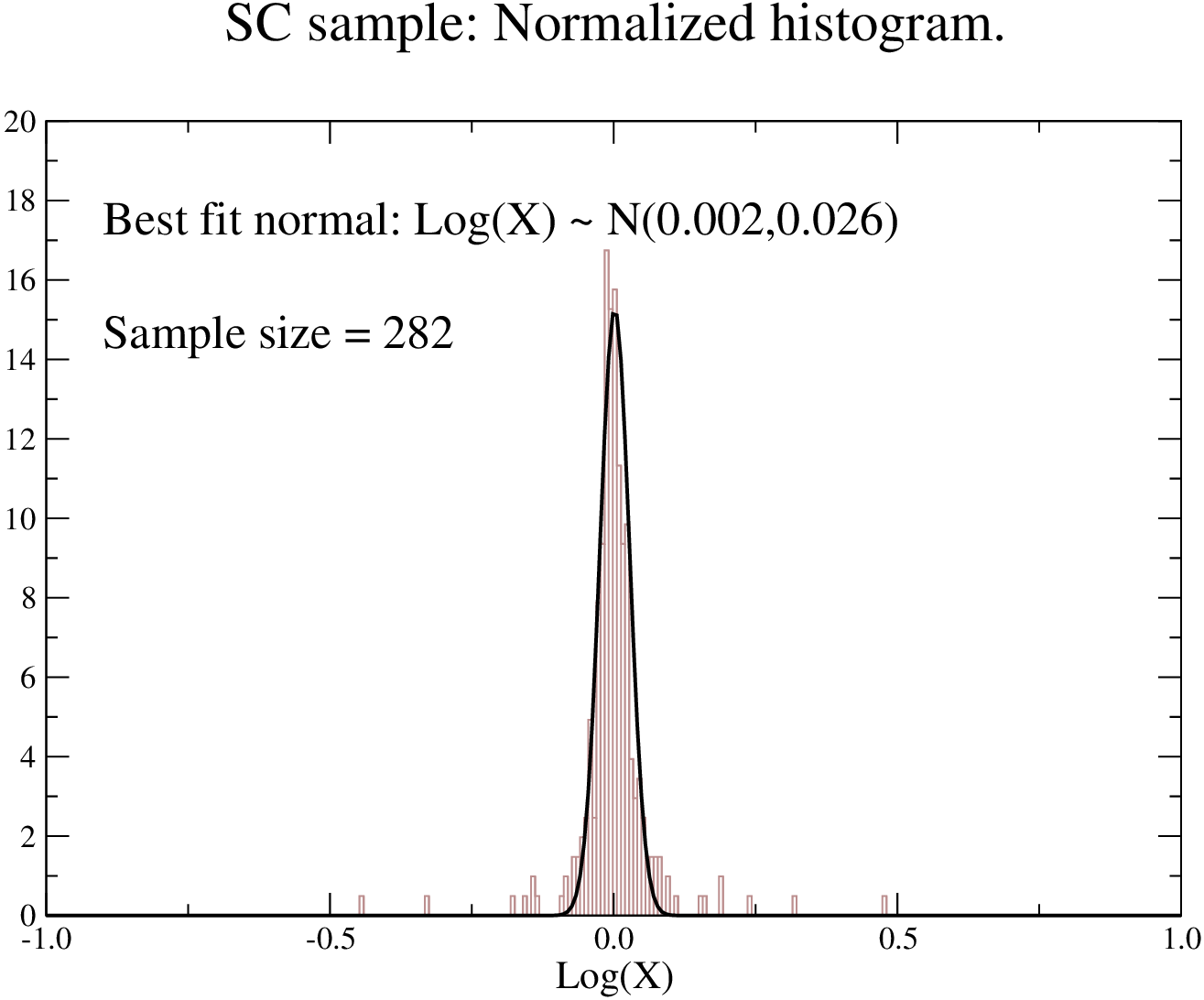}\\

\caption{\label{fig2} Distributions of $\log X$ for each of four large ORC
samples.}

\end{figure}

\subsection{The analysis}

Any rotation curve can be expressed as\begin{equation}
\left(\frac{V}{V_{0}}\right)=\left(\frac{R}{R_{0}}\right)^{\alpha}+g\left(\frac{R}{R_{0}}\right)\label{eqn8A}\end{equation}
for arbitrary $\left(R_{0}>0,\, V_{0}>0,\,\alpha\right)$ and a suitably
chosen function, $g\left(R/R_{0}\right)$. From this there follows
immediately\begin{eqnarray}
\left(\frac{V}{V_{0}}\right) & = & \left(\frac{R}{R_{0}}\right)^{\alpha}\left[1+\left(\frac{R_{0}}{R}\right)^{\alpha}g\left(\frac{R}{R_{0}}\right)\right]\equiv\left(\frac{R}{R_{0}}\right)^{\alpha}X\nonumber \\
 & \downarrow\nonumber \\
\log\left(\frac{V}{V_{0}}\right) & = & \alpha\log\left(\frac{R}{R_{0}}\right)+\log X\,.\label{eqn11}\end{eqnarray}
In the following, we show that, on the basis of the four large samples,
$X=1$ in a statistical sense. The analysis proceeds as follows:
\begin{itemize}
\item choose the scaling functions relating $\left(V_{0},R_{0}\right)$
to luminosity properties, as one of (\ref{eqn9A}), (\ref{eqn9B}),
(\ref{eqn9C}) or (\ref{eqn9D}) depending on the ORC sample being
considered; 
\item with the chosen scaling functions, perform a linear regression of
$\log\left(V/V_{0}\right)$ on $\log\left(R/R_{0}\right)$ as described
in \S\ref{Extraction};
\item we now have a linear model for (\ref{eqn11}) for the chosen ORC where
the gradient gives an estimate of $\alpha$ and the zero point represents
the $\log X$ term. 
\item the normalized histograms of the $\log X$ distributions, together
with the best fitting pdf, is given in figure 2. For each of the four
samples, the best pdf fits are given in the table:
\end{itemize}
~~~~~~~~~~~~~~~~~~~~~~~~~~~\begin{tabular}{|r|c|}
\hline 
Sample & $N(\bar{x},\sigma)$ fitted to $\log X$\tabularnewline
\hline 
MFB & $N(-0.001,0.043)$\tabularnewline
\hline 
MF & $N(+0.001,0.039)$\tabularnewline
\hline 
SC & $N(+0.002,0.026)$\tabularnewline
\hline
DGHU & $N(+0.001,0.033)$\tabularnewline
\hline
\end{tabular}
\begin{itemize}
\item It follows from the latter table and (\ref{eqn11}) that $X=1$ in
the statistical sense, and at the level of near statistical certainty.
In other words, there is absolutely no evidence to suggest that the
power-law hypothesis for rotation velocities in the annular disc is
not supported and that, for all practical purposes, we have \[
\left(\frac{V}{V_{0}}\right)\approx\left(\frac{R}{R_{0}}\right)^{\alpha},\,\,\, R\in{\rm annular\, disc}\,\]
 for the annular disc in each of the four samples analysed. 
\end{itemize}

\section{Conclusions}

It has been shown, at the level of statistical certainty, that:
\begin{itemize}
\item the optical disc of spiral galaxies consists of two quite distinct
dynamical regions comprising an interior central sub-disc and a surrounding
annular disc with an objectively defined dynamical transition boundary
separating the two regions;
\item that a scaling law, similar to the classical Tully-Fisher scaling
law, applies on the \emph{interior }boundary of the annular disc.
\end{itemize}
Furthermore, the analysis has provided a refinement of the Tully-Fisher
scaling law, involving an additional parameter, which renders it insensitive
to the precise definition of (optical) rotation velocity employed.

\appendix

\section{The samples of MF and SC and $\alpha\log V_{rot}$ }

In the following, we describe the analyses corresponding to that of
\S\ref{sec:4} applied to the samples of Mathewson \& Ford, and Courteaux.
For each sample, we use two definitions of $V_{rot}$:
\begin{itemize}
\item $V_{rot}=V_{author}$;
\item $V_{rot}=V_{opt}$ defined as the rotational velocity estimated at
$R_{83}$.
\end{itemize}
For each sample, we find that the use of $V_{opt}$ in the classical
Tully-Fisher scaling law gives very significantly worse results that
the use of $V_{author}$. However, we then find that performance of
the enhanced Tully-Fisher law (with $\alpha\log V_{rot}$) is insensitive
to the precise definition used of $V_{rot}$ with, for example, $V_{opt}$
being comparable to $V_{author}$ for overall effectiveness.

\subsection{Sample of MFB using $V_{rot}=V_{hist}$}

It is clear from their paper that MFB estimated $V_{rot}$ for their
Tully-Fisher work using a version of the histogram method, but it
is not clear, either from their words or their data, exactly how they
implimented the method in practice. In the following, we calibrate
the basic Tully-Fisher law using MFB's estimates for $V_{rot}$ and
absolute magnitudes, $M_{H}$ say, calculated from Hubble distances
with $H=85km/sec/kpc$.

\subsubsection*{The classical Tully-Fisher scaling law with $V_{rot}=V_{hist}$}

We find, for the classical Tully-Fisher model,\begin{eqnarray}
M_{H}\approx M_{TF} & = & (-6.72\pm0.21)+(-6.54\pm0.10)\log V_{rot};\nonumber \\
 &  & (n=837,\,\,\, t_{grad}=-69)\label{eqn5}\\
 &  & Adj\, R^{2}=85\%,\,\,\, RSE=0.45\nonumber \end{eqnarray}
after removing 26 of the most extreme outliers. The quoted $t$-statistic
refers to the gradient estimate. We note that the gradient lies well
within the accepted envelope.

\subsubsection*{The enhanced Tully-Fisher scaling law with $V_{rot}=V_{hist}$}

We find, for the enhanced model, using exactly the same reduced data
set,\begin{eqnarray}
M_{H}\approx M_{TF} & = & (-7.47\pm0.23)+\left[\left(-6.28\pm0.10\right)+\left(0.22\pm0.03\right)\alpha\right]\log V_{rot};\nonumber \\
 &  & (n=837,\,\,\, t_{grad}=-62,\,\,\, t_{\alpha}=6.5)\label{eqn6}\\
 &  & Adj\, R^{2}=86\%,\,\,\, RSE=0.44\nonumber \end{eqnarray}
so that, according to the $t$-statistic for the gradient of the $\alpha\log V_{rot}$
component ($t_{\alpha}=6.5$), this component appears to be present
in the model in a highly significant way. However, each model explains
about $85\%$ of the variation in the data so that the inclusion of
$\alpha$ does not make much difference to model's predictive power
in this case.

\subsection{Sample of MFB using $V_{rot}=V_{opt}$\label{sub:Sample-of-Mathewson,}}

In the following we show that, for the MFB sample, $V_{rot}=V_{opt}$
performs poorly compared to MFB's own determination, $V_{rot}=V_{hist}$,
but that when $\alpha$ is included as a predictor, the enhanced model
(using $V_{rot}=V_{opt}$) is directly comparable to models (\ref{eqn5})
and (\ref{eqn6}) above.

\subsubsection*{The classical Tully-Fisher scaling law with $V_{rot}=V_{opt}$}

We find\begin{eqnarray*}
M_{H}\approx M_{TF} & = & (-7.70\pm0.28)+(-5.98\pm0.13)\log V_{opt};\\
 &  & (n=837,\,\,\, t_{grad}=-47)\\
 &  & Adj\, R^{2}=73\%,\,\,\, RSE=0.60\end{eqnarray*}
after removing twelve of the most extreme outliers. The quoted $t$-statistic
refers to the gradient estimate. Note that this model explains only
about 73\% of the variation in the data compared to about 85\% for
model (\ref{eqn5}) above.

\subsubsection*{The enhanced Tully-Fisher scaling law with $V_{rot}=V_{opt}$}

Using exactly the same reduced data set as above, we find\begin{eqnarray}
M_{H}\approx M_{TF} & = & (-8.88\pm0.22)+\left[\left(-5.73\pm0.10\right)+\left(0.78\pm0.03\right)\alpha\right]\log V_{opt};\nonumber \\
 &  & (n=837,\,\,\, t_{grad}=-57,\,\,\, t_{\alpha}=26)\label{eqn6a}\\
 &  & Adj\, R^{2}=85\%,\,\,\, RSE=0.45.\nonumber \end{eqnarray}
It is clear that the $\alpha\log V_{opt}$ component ($t_{\alpha}=26$)
is very powerfully present in the model. Furthermore, this model now
explains about $85\%$ of the variation in the data (up from 73\%)
and so is directly comparable with models (\ref{eqn5}) and (\ref{eqn6})
which both use MFB's own $V_{rot}=V_{hist}$ determinations.

\subsection{Sample of Dale et al using $V_{rot}=V_{hist}$}

The primary conclusion from the foregoing analysis of MFB data is
that the inclusion of the predictor $\alpha\ln V_{rot}$ significantly
enhances the performance of the Tully-Fisher scaling law. However,
it transpires that the degree of this enhancement depends on the precise
definition of $V_{rot}$ employed, as we now demonstrate via the analysis
of the DGHU data.

\subsubsection*{The classical Tully-Fisher scaling law with $V_{rot}=V_{hist}$}

If we calibrate the basic Tully-Fisher law using Dale's own estimates
($0.5\left(V_{90\%}-V_{10\%}\right)$) for $V_{rot}$ and absolute
magnitudes calculated from Hubble distances with $H=85km/sec/kpc$
we find, for the TF classical model,\begin{eqnarray*}
M_{TF} & = & (-7.18\pm0.28)+(-6.48\pm0.13)\log V_{rot};\,\,\,(n=478,\,\,\, t=-52)\\
 &  & Adj\, R^{2}=85\%,\,\,\, RSE=0.40\end{eqnarray*}
after removing eighteen of the most extreme outliers. The quoted $t$-statistic
refers to the gradient estimate. We note that the gradient lies well
within the accepted envelope.

\subsubsection*{The enhanced Tully-Fisher scaling law with $V_{rot}=V_{hist}$}

We find, for the enhanced model, using exactly the same reduced data
set:\begin{eqnarray*}
M_{TF} & = & (-7.39\pm0.31)+\left[\left(-6.41\pm0.14\right)+\left(0.07\pm0.04\right)\alpha\right]\log V_{rot};\\
 &  & (n=478,\,\,\, t_{grad}=-47,\,\,\, t_{\alpha}=1.5)\\
 &  & Adj\, R^{2}=85\%,\,\,\, RSE=0.40\end{eqnarray*}
so that, according to the $t$-statistic for the gradient of the $\alpha\log V_{rot}$
component ($t_{\alpha}=1.5$), this component appears to be \emph{not
significantly} present in the model.

\subsection{Sample of Dale using $V_{rot}=V_{opt}$}

To emphasize the point that it is Dale's definition of $V_{rot}$
which makes $\alpha\log V_{rot}$ insignificant, rather than some
property of the sample, we repeat the analysis of DGHU data, but using
$V_{rot}\equiv V_{opt}$.

\subsubsection*{The classical Tully-Fisher scaling law with $V_{rot}=V_{opt}$}

We find, for the classical TF model using $V_{opt}=V(R_{83})$,\begin{eqnarray*}
M_{TF} & = & (-7.54\pm0.38)+(-6.22\pm0.17)\log V_{opt};\,\,\,(n=477,\,\,\, t=-37)\\
 &  & Adj\, R^{2}=74\%,\,\,\, RSE=0.53\end{eqnarray*}
after removing eighteen of the most extreme outliers. The quoted $t$-statistic
refers to the gradient estimate. We note that this model is considerably
less effective in explaining the data than when $V_{rot}=V_{hist}$
is used.

\subsubsection*{The enhanced Tully-Fisher scaling law with $V_{rot}=V_{opt}$}

For the enhanced model, we find\begin{eqnarray*}
M_{TF} & = & (-8.40\pm0.32)+\left[\left(-6.01\pm0.14\right)+\left(0.61\pm0.04\right)\alpha\right]\log V_{opt};\\
 &  & (n=478,\,\,\, t_{grad}=-43,\,\,\, t_{\alpha}=15)\\
 &  & Adj\, R^{2}=82\%,\,\,\, RSE=0.44\end{eqnarray*}
so that, according to the $t$-statistic for the gradient of the $\alpha\log V_{rot}$
component ($t_{\alpha}=15$), this component appears to be present
in the model in an extremely significant way. Furthermore, the inclusion
of the predictor $\alpha\log V_{rot}$ has sufficiently enhanced the
predictive power of the model so that it is comparable with the model
which uses $V_{rot}=V_{hist}$.

\subsection{Sample of MF using $V_{rot}=V_{hist}$}

\subsubsection*{The classical Tully-Fisher scaling law with $V_{rot}=V_{hist}$}

If we calibrate the basic Tully-Fisher law using MF's own estimates
for $V_{rot}$ and absolute magnitudes calculated from Hubble distances
with $H=85km/sec/kpc$ we find, for the classical TF model,\begin{eqnarray}
M_{TF} & = & (-9.23\pm0.21)+(-5.50\pm0.10)\log V_{rot};\,\,\,(n=1057,\,\,\, t=-58)\label{eqn7}\\
 &  & Adj\, R^{2}=76\%,\,\,\, RSE=0.46\nonumber \end{eqnarray}
after removing 28 of the most extreme outliers. The quoted $t$-statistic
refers to the gradient estimate. We note that the gradient is on the
low side of expected values.

\subsubsection*{The enhanced Tully-Fisher scaling law with $V_{rot}=V_{hist}$}

We find, for the enhanced model, using exactly the same reduced data
set,\begin{eqnarray*}
M_{TF} & = & (-9.88\pm0.23)+\left[\left(-5.27\pm0.10\right)+\left(0.21\pm0.03\right)\alpha\right]\log V_{rot};\\
 &  & (n=1057,\,\,\, t_{grad}=-53,\,\,\, t_{\alpha}=6.6)\\
 &  & Adj\, R^{2}=77\%,\,\,\, RSE=0.46\end{eqnarray*}
so that, according to the $t$-statistic for the gradient of the $\alpha\log V_{rot}$
component ($t_{\alpha}=6.6$), this component appears to be present
in the model in a highly significant way. \\
\\
We see that the inclusion of the predictor $\alpha\ln V_{rot}$
gives a significant, but small, improvement in performance.

\subsection{Sample of MF using $V_{rot}=V_{opt}$}

\subsubsection*{The classical Tully-Fisher scaling law with $V_{rot}=V_{opt}$}

If we calibrate the basic Tully-Fisher law using $V_{opt}\equiv V(R_{83})$
we find, for the classical TF model,\begin{eqnarray*}
M_{TF} & = & (-9.83\pm0.25)+(-5.11\pm0.11)\log V_{opt};\,\,\,(n=1053,\,\,\, t=-47)\\
 &  & Adj\, R^{2}=67\%,\,\,\, RSE=0.54\end{eqnarray*}
after removing 32 of the most extreme outliers. We see that the model
explains about 67\% of the variation in the data, compared with about
76\% for model (\ref{eqn7}). It is thus very much less effective.

\subsubsection*{The enhanced Tully-Fisher scaling law with $V_{rot}=V_{opt}$}

However, if we now consider the enhanced model, we find\begin{eqnarray*}
M_{TF} & = & (-10.49\pm0.20)+\left[\left(-5.04\pm0.09\right)+\left(0.61\pm0.03\right)\alpha\right]\log V_{opt};\\
 &  & (n=1053,\,\,\, t_{grad}=-56,\,\,\, t_{\alpha}=23)\\
 &  & Adj\, R^{2}=78\%,\,\,\, RSE=0.44\end{eqnarray*}
so that, according to the $t$-statistic for the gradient of the $\alpha\log V_{rot}$
component ($t_{\alpha}=23$), this component is very powerfully present
in the model. Furthermore, the model now explains about 78\% of the
variation in the data and is therefore significantly \emph{better}
than model (\ref{eqn7}) which uses MF's own estimate of $V_{rot}$.

\subsection{Sample of SC using $V_{rot}=V_{model}\equiv V_{max}$}

Courteaux's paper is primarily a study of various methods of estimating
$V_{rot}$ for Tully-Fisher studies and his estimator $V_{max}$ is
derived from as the peak velocity achieved by a particular phenomenological
model.

\subsubsection*{The classical Tully-Fisher scaling law with $V_{rot}=V_{max}$}

If we calibrate the basic Tully-Fisher law using Courteaux's $V_{max}$
and absolute magnitudes calculated from Hubble distances with $H=85km/sec/kpc$
we find, for the classical TF model,\begin{eqnarray*}
M_{TF} & = & (-5.53\pm0.50)+(-6.66\pm0.22)\log V_{rot};\,\,\,(n=254,\,\,\, t=-30)\\
 &  & Adj\, R^{2}=78\%,\,\,\, RSE=0.39\end{eqnarray*}
after removing twenty of the most extreme outliers. The quoted $t$-statistic
refers to the gradient estimate. We note that the gradient lies well
within the accepted envelope.

\subsubsection*{The enhanced Tully-Fisher scaling law with $V_{rot}=V_{max}$}

We find, for the enhanced model, using exactly the same data set\begin{eqnarray*}
M_{TF} & = & (-6.85\pm0.52)+\left[\left(-6.17\pm0.22\right)+\left(0.36\pm0.06\right)\alpha\right]\log V_{rot};\\
 &  & (n=254,\,\,\, t_{grad}=-28,\,\,\, t_{\alpha}=5.9)\\
 &  & Adj\, R^{2}=81\%,\,\,\, RSE=0.36\end{eqnarray*}
so that, according to the $t$-statistic for the gradient of the $\alpha\log V_{rot}$
component ($t_{\alpha}=6.3$), this component appears to be present
in the model in a highly significant way.

\subsection{Sample of SC using $V_{rot}=V_{model}\equiv V_{2.2}$}

Courteaux's paper is primarily a study of various methods of estimating
$V_{rot}$ for Tully-Fisher studies and his estimator $V_{2.2}$ is
the model velocity at $2.15\times$scale lengths.

\subsubsection*{The classical Tully-Fisher scaling law with $V_{rot}=V_{2.2}$}

If we calibrate the basic Tully-Fisher law using Courteaux's $V_{2.2}$
and absolute magnitudes calculated from Hubble distances with $H=85km/sec/kpc$
we find, for the classical TF model,\begin{eqnarray*}
M_{TF} & = & (-6.43\pm0.41)+(-6.33\pm0.18)\log V_{rot};\,\,\,(n=262,\,\,\, t=-34)\\
 &  & Adj\, R^{2}=82\%,\,\,\, RSE=0.36\end{eqnarray*}
after removing eighteen of the most extreme outliers. The quoted $t$-statistic
refers to the gradient estimate. We note that the gradient lies well
within the accepted envelope.

\subsubsection*{The enhanced Tully-Fisher scaling law with $V_{rot}=V_{2.2}$}

We find, for the enhanced model, using exactly the same data set\begin{eqnarray*}
M_{TF} & = & (-6.83\pm0.47)+\left[\left(-6.18\pm0.20\right)+\left(0.10\pm0.06\right)\alpha\right]\log V_{rot};\\
 &  & (n=262,\,\,\, t_{grad}=-31,\,\,\, t_{\alpha}=1.8)\\
 &  & Adj\, R^{2}=82\%,\,\,\, RSE=0.36\end{eqnarray*}
so that, according to the $t$-statistic for the gradient of the $\alpha\log V_{rot}$
component ($t_{\alpha}=6.3$), this component appears to be present
in the model in a highly significant way.

\subsection{Sample of SC using $V_{rot}=V_{opt}$}

\subsubsection*{The classical Tully-Fisher scaling law with $V_{rot}=V_{opt}$}

If we calibrate the basic Tully-Fisher law using Courteaux's $V_{opt}$
and absolute magnitudes calculated from Hubble distances with $H=85km/sec/kpc$
we find, for the classical TF model,\begin{eqnarray*}
M_{TF} & = & (-6.59\pm0.64)+(-6.12\pm0.28)\log V_{opt};\,\,\,(n=254,\,\,\, t=-22)\\
 &  & Adj\, R^{2}=65\%,\,\,\, RSE=0.49\end{eqnarray*}
after removing 21 of the most extreme outliers. The quoted $t$-statistic
refers to the gradient estimate. We note that the gradient lies well
within the accepted envelope.

\subsubsection*{The enhanced Tully-Fisher scaling law with $V_{rot}=V_{opt}$}

For the enhanced model, we find\begin{eqnarray*}
M_{TF} & = & (-7.84\pm0.51)+\left[\left(-5.76\pm0.22\right)+\left(0.75\pm0.06\right)\alpha\right]\log V_{opt};\\
 &  & (n=254,\,\,\, t_{grad}=-26,\,\,\, t_{\alpha}=13)\\
 &  & Adj\, R^{2}=79\%,\,\,\, RSE=0.38\end{eqnarray*}
so that, according to the $t$-statistic for the gradient of the $\alpha\log V_{rot}$
component ($t_{\alpha}=134$), this component is present in the model
in an extremely significant way.

\end{document}